\documentclass[a4paper,11pt]{article}
\pdfoutput=1 

\usepackage{amssymb,amsmath}
\usepackage{verbatim,xspace}
\usepackage[version=3]{mhchem}

\usepackage{jcappub} 

\usepackage[T1]{fontenc} 

\newcommand{\ro}[1]{\ensuremath{\textrm{#1}}}
\newcommand{\ten}[1]{\ensuremath{\times 10^{#1}}}

\newcommand{\Msol}{\ensuremath{M_{\odot}}\xspace}

\newcommand{\df}{\ensuremath{~ \ro{d}}}
\newcommand{\dd}{\ensuremath{\ro{d}}}

\newcommand{\sv}{\ensuremath{~\langle \sigma v \rangle}\xspace}
\newcommand{\HeW}{\ensuremath{\ce{^{2}He}}\xspace}
\newcommand{\HeT}{\ensuremath{\ce{^{3}He}}\xspace}
\newcommand{\HeF}{\ensuremath{\ce{^{4}He}}\xspace}
\newcommand{\CTw}{\ensuremath{\ce{^{12}C}}\xspace}
\newcommand{\Kp}{\ensuremath{\mathcal{K}_{pep}}\xspace}

\newcommand{\mPr}{\ensuremath{m_\ro{p}}\xspace}

\newcommand{\mR}{\ensuremath{m_\ro{R}}\xspace}
\newcommand{\mion}{\ensuremath{m_\ro{i}}\xspace}
\newcommand{\cC}{\ensuremath{\mathcal{C}}\xspace}
\newcommand{\Deu}{\ensuremath{[\ro{D}/\ro{H}]}\xspace}

\title{Producing the Deuteron in Stars: Anthropic Limits on Fundamental Constants}

\author[a]{Luke A. Barnes and Geraint F. Lewis}

\affiliation[a]{Sydney Institute for Astronomy   \\
School of Physics, A28   \\
The University of Sydney   \\
NSW 2006, Australia}

\emailAdd{luke.barnes@sydney.edu.au}

\abstract{Stellar nucleosynthesis proceeds via the deuteron (D), but only a small change in the fundamental constants of nature is required to unbind it. Here, we investigate the effect of altering the binding energy of the deuteron on proton burning in stars. We find that the most definitive boundary in parameter space that divides probably life-permitting universes from probably life-prohibiting ones is between a bound and unbound deuteron. Due to neutrino losses, a ball of gas will undergo rapid cooling or stabilization by electron degeneracy pressure before it can form a stable, nuclear reaction-sustaining star. We also consider a less-bound deuteron, which changes the energetics of the $pp$ and $pep$ reactions. The transition to endothermic $pp$ and $pep$ reactions, and the resulting beta-decay instability of the deuteron, do not seem to present catastrophic problems for life.}

\begin{document}
\maketitle
\flushbottom

\section{Introduction}
\label{sec:intro}

The free parameters of physics --- fundamental constants and cosmological parameters --- seem to be finely-tuned for life. Small changes in their values can suppress or erase the complexity upon which physical life as we know it, or can imagine it, depends. The scientific literature on the fine-tuning of the universe for life has been reviewed in several articles \cite{Davies1983,Barrow1986,Hogan2000,Barnes2012,Schellekens2013,Lewis2016}.

In a previous paper, we considered the potentially life-prohibiting effects of changing the fundamental constants of nature so that the diproton is bound \cite[][hereafter B15]{Barnes2015}. While it had been predicted that diproton-burning stars would be explosive, we found that they are remarkably similar to stars in our universe; they have similar luminosities at a given mass, and smaller stars are possible because nuclear ignition temperatures are lower. This understanding of strong-burning stars places limits on the region of parameter space in which stars can burn at all.

Here, we consider a different scenario: rather than binding all two-nucleon nuclei, we gradually unbind them all. Barrow and Tipler \cite[][p. 322]{Barrow1986} argue that ``if the nuclear force were a little weaker the deuteron would be unbound with other adverse consequences for the nucleosynthesis of biological elements because a key link in the chain of nucleosynthesis would be removed. Elements heavier than hydrogen would not form.'' Davies \cite{Davies1983} says ``Without deuterium, the main nuclear reaction chain used by the sun could not proceed. It is doubtful if stable, long-lived stars could exist at all.''

The dependence of the properties of the deuteron (mass, binding energy) on the fundamental constants of nature has been calculated using a number of approaches  \cite{Davies1972,Agrawal1998a,Agrawal1998b,Hogan2000,
Beane2002,Beane2003,Epelbaum2005,Hogan2006,Flambaum2007,
Golowich2008,Jaffe2009,Ekstrom2010,Uzan2011,Ali2013}. In particular, a number of authors have studied deuteron in order to calculate the dependence of big bang nucleosynthesis on the light quark masses. This, in turn, places observational limits on possible variations of these quantities in the early universe \cite{Bedaque2011,Coc2012,Berengut2013,Carrillo2013}. The dependence of the deuteron binding energy ($B_D$) on the sum of the light quark masses varies between these papers; following Barr \& Khan \cite{Barr2007}, a typical conservative parameterization of this relationship is,
\begin{equation} \label{eq:BDquark}
B_D = 2.2 \ro{MeV} - 1.3 \ro{MeV} \left(\frac{m_{u} + m_{d}}{m_{u,0} + m_{d,0}} - 1 \right) ~,
\end{equation}
where $m_{u,0} = 2.3$ MeV and  $m_{d,0} = 4.8$ MeV are the up and down quark masses in our universe \cite{Olive2014}. Equation \ref{eq:BDquark} implies that the deuteron is unbound $(B_D < 0)$ when $m_{u} + m_{d} > 19.1$ MeV. Note, however, that the theoretical situation is not fully settled. \cite{Beane2002,Beane2003} noted that in 2002-3, Lattice QCD (LQCD) calculations had not conclusively shown how $B_D$ depends on the light quark masses, or even the sign of the dependence. This situation persists; in 2016, Baru et al. \cite{Baru2015,Baru2016} note that the  ``current situation with the lattice-QCD results obtained by different groups is not completely clear''.

In light of this uncertainty, we will investigate the binding of the deuteron by changing the strength of the nuclear strong force. A $\sim 5-8$\% decrease in the strength of the strong force (specifically, the attractive and repulsive terms in the potential) will unbind the deuteron \cite{Davies1972,Davies1983,Pochet1991,Cohen2008,Golowich2008}. If the binding energy of the deuteron nucleus ($B_D$) were negative, then a deuteron would break into its component nucleons ($D \rightarrow p + n$) on extremely short time scales ($\sim 10^{-23}$ seconds), which is approximately the time taken for the nucleons to travel across the nuclear potential well. 

In universes in which the deuteron is bound, altering its binding energy can still effect the energetics of stellar burning. The relevant nuclear processes in stars are as follows:
\begin{itemize}
\item The $pp$ reaction $p + p \rightarrow D + e^+ + \nu_e$ releases energy,
\begin{equation} \label{eq:Qpp}
Q_{pp} = B_D - (m_n - m_p) -  m_e  ~,
\end{equation}
where $m_e$, $m_n$ and $m_p$ are the electron, neutron and proton mass respectively; we will assume throughout that the neutrino is effectively massless. In our universe, $Q_{pp} = 0.420$ MeV. Altering the binding energy of the deuteron can result in an endothermic $pp$ reaction $(Q_{pp} < 0)$.
\item The $pep$ reaction $p + e^- + p \rightarrow D + \nu$, being a three body reaction, has a lower reaction rate in stars in our universe and is thus a minor contributor to stellar burning. It releases energy,
\begin{equation} \label{eq:Qpep}
Q_{pep} = B_D  - (m_n - m_p) +  m_e ~.
\end{equation}
In our universe, $Q_{pep} = 1.442$ MeV. 
\item If the deuteron is sufficiently loosely bound, then it can be unstable to beta decay, $D \rightarrow p + p + e^- + \bar{\nu}_e$. This decay will be energetically possible if $Q^\beta_D \equiv (m_n - m_p) - m_e - B_D > 0$. Note that, because we assume that the neutrino has negligible mass, $Q_{pep} = -Q^\beta_D$. In our universe, $Q^\beta_D = -1.442$ MeV. 
\end{itemize}

\begin{figure*} \centering
	\begin{minipage}{0.35\textwidth}
		\includegraphics[width=\textwidth]{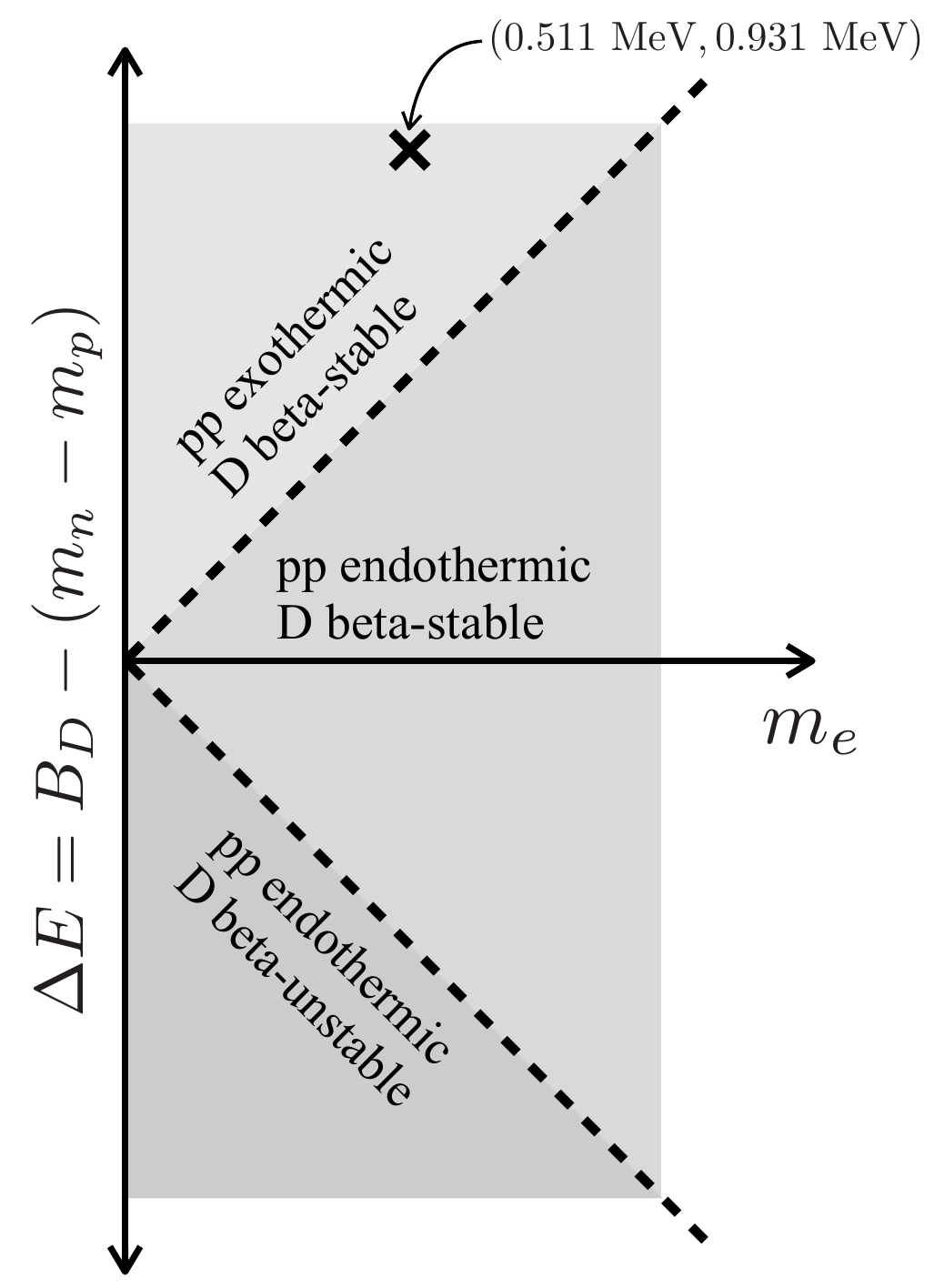}
	\end{minipage}
	\begin{minipage}{0.45\textwidth}
		\caption{The regions of interest for a bound deuteron. We define $\Delta E = B_D - (m_n - m_p)$; then $Q_{pp} = \Delta E - m_e$ and $Q_{pep}  = -Q^\beta_D = \Delta E + m_e$. The top region ($\Delta E > m_e$) is where our universe is found ($m_e = 0.511$ MeV, $\Delta E = 0.931$ MeV), in which the $pp$ and $pep$ reactions are exothermic (D is $\beta$-decay stable). In the middle region ($-m_e < \Delta E < m_e$) the $pp$ reaction is endothermic and $pep$ is exothermic (D is $\beta$-decay stable). In the bottom region ($\Delta E > m_e$), the $pp$ and $pep$ reactions are endothermic (D is $\beta$-decay unstable).}
	\end{minipage}
\label{fig:Dme}
\end{figure*}

We summarize the relevant regimes in Figure \ref{fig:Dme}. By defining $\Delta E = B_D - (m_n - m_p)$, we can write $Q_{pp} = \Delta E - m_e$ and $Q_{pep}  = -Q^\beta_D = \Delta E + m_e$. The top region ($\Delta E > m_e$) is where our universe is found ($m_e = 0.511$ MeV, $\Delta E = 0.931$ MeV), in which the $pp$ and $pep$ reactions are exothermic (D is $\beta$-decay stable). In the middle region ($-m_e < \Delta E < m_e$) the $pp$ reaction is endothermic and $pep$ is exothermic (D is $\beta$-decay stable). In the bottom region ($\Delta E > m_e$), the $pp$ and $pep$ reactions are endothermic (D is $\beta$-decay unstable). Note that there is no region in which D is $\beta$-decay unstable and $pp$ is exothermic. Thus, we cannot consider the effect of D $\beta$-decay without first considering the effect of an endothermic $pp$ reaction.

Since the positron created in the $pp$ reaction will rapidly annihilate with an electron, releasing $2 m_e c^2$ of energy, the actual energy released per $pp$ reaction is the same as per $pep$ reaction. Also, even if $D$ is produced by an endothermic reaction, if it quickly reacts with protons to produce \HeT and \HeF via $p + D \rightarrow \HeT $ and $\HeT  + \HeT \rightarrow \HeF + 2p$, then the entire chain could still be exothermic, especially because the binding energy per nucleon is much higher for \HeF than $D$ (7.1 MeV vs. 1.1 MeV). In other words, nuclear burning to \HeF can produce energy, even when burning to $D$ consumes energy.

However, endothermic reaction steps can still dramatically affect stellar burning by influencing the reaction rate. The required energy ($-|Q_{pp}|$) must be supplied by the kinetic energy of the reactants, and so the reaction rate will be exponentially suppressed at low temperatures.

In this paper we will consider the properties of stars in universes in which the nuclear production of deuterium is different.  In Section \ref{S:StarModel}, we review the stellar model first presented in Adams \cite{Adams2008}, and which was used in B15 to model diproton-burning stars. We consider the alterations and extensions required to model stars in which the deuteron is unstable. In Section \ref{S:UnstableD}, we consider the effects of an unstable deuteron, varying its lifetime. Returning to the case of a bound deuteron, in Section \ref{S:endopp}, we consider the effect of an endothermic $pp$ reaction. We consider the effects of varying the electron mass in Section \ref{S:varyme}. Finally, in Section \ref{S:starparam} we summarize our results.

\section{Stellar Model} \label{S:StarModel}
\subsection{Extending the Stellar Model of Adams (2008)}
Our stellar model follows the formalism of Adams (2008) \cite{Adams2008}. We will rely on the summary of Adams's model in B15; here we will focus on the extensions to the model required by an unstable deuteron.

A system of four coupled differential equations describes the structure of the star: force balance (hydrostatic equilibrium), conservation of mass, heat transport and energy generation. From hydrostatic equilibrium, given an equation of state $P = K \rho^{\Gamma}$, we can derive the dimensionless temperature and density profile for the star, given the constraint of its total mass $M_*$. To calculate the central temperature $(T_c)$ of the star, we need to consider the rate of nuclear energy production. Adams's model uses a simple parameterization of a two-body reaction rate per unit volume, as a function of radius,
\begin{equation}
\epsilon(r) = \cC \rho^2 \Theta^2 \exp(-3\Theta) ~,
\end{equation}
where $\cC = 8 Q S_0 / (\sqrt{3} \pi \alpha m_1 m_2 Z_1 Z_2 \mR c)$ is a constant (in a given universe) that summarizes the relevant nuclear physics, $\Theta =  (E_G/4k_B T)^{1/3}$, $E_G = (\pi \alpha Z_1 Z_2)^2 2 \mR c^2$ is the Gamow energy for the reaction, $Q$ is the average energy released per reaction, $S_0$ is the so-called astrophysical $S$-factor, $m_1$ and $m_2$ ($Z_1$ and $Z_2$) are the masses (electric charges) of the reacting nuclei, $\mR$ is their reduced mass, and $\alpha$ is the fine-structure constant.

However, as the lifetime of the deuteron decreases (while \HeT and \HeF remain bound), the reaction becomes effectively three-body. Additionally, if the $pp$ reaction is endothermic, then its suppression may lead to the increased importance of the three-body $pep$ reaction. The overall proton-burning pathway is,
\begin{align}
p + p 			 &\rightarrow D + e^+ + \nu_e  		\label{eq:React1}   				\\
p + p + e^-	 &\rightarrow D + \nu_e  				\label{eq:React1e}   					\\
D 					 &\rightarrow p + p + e^- + \bar{\nu}_e  \label{eq:React2b}   	\\
D + p 			 &\rightarrow \HeT  						\label{eq:React3}   				\\
\HeT + \HeT &\rightarrow \HeF + 2p ~,  			\label{eq:React4}
\end{align}
If the deuteron is unbound ($B_D < 0$), then as well as beta decay (Reaction \ref{eq:React2b}), the deuteron can disintegrate directly into a proton and a neutron. This creates free neutrons, which gives another set of reactions that we must take into account.
\begin{align}
D 					 &\leftrightarrow p + n  						\label{eq:React2d}   	\\
n 					 &\rightarrow p + e^- + \bar{\nu}_e  \label{eq:React2n} \\
\HeW + n		 &\rightarrow \HeT  							\label{eq:React5}   
\end{align}
Reaction \ref{eq:React5}, given the very short lifetime of the unstable diproton, is essentially the second stage of a three-body ($p+p+n$) reaction. Its effect is negligible relative to the $D + p$ reaction: to produce to a stable \HeT nucleus, the neutron combines with a diproton, which has a much shorter lifetime than the deuteron for the cases considered here. While we ignore this reaction in the equations below, we have kept the reaction in our code to confirm that it is negligible, assuming its rate \sv to be comparable to that of the $p + n$ reaction. Nevertheless, the $p+n$ reaction \eqref{eq:React2d} is an important factor in determining the equilibrium abundance of deuterium. 

For the species intermediate between protons and \HeF, their abundance will quickly approach an equilibrium value:
\begin{align} \label{eq:dneqzero}
\frac{\df n_D}{\df t} &= \frac{1}{2} n_p^2 \sv_{pp} +  n_p^2 n_e \Kp + n_p n_n \sv_{pn} - n_p n_D \sv_{Dp} - (\lambda_D + \lambda^\beta_D) n_D  = 0 ~, \\
\frac{\df n_n}{\df t} &= \lambda_D n_D - n_p n_n \sv_{pn} - \lambda_n n_n = 0 ~,
\end{align}
where $n_e$, $n_p$, $n_n$ and $n_D$ are the number density of electrons, protons, neutrons and deuterons respectively; $\sv_{pp}$, $\sv_{pn}$ and $\sv_{Dp}$ are the thermally-averaged reaction rates per particle pair for the $pp$, $pn$ and $Dp$ reaction respectively; \Kp  is the corresponding reaction rate parameter for the three-body $pep$ reaction \cite{Irgaziev2014}; $\lambda_n = 7.86 \ten{-4}$ s is the neutron decay constant; $\lambda_D$ is the decay constant for deuterium unbinding, and $\lambda^\beta_D$ is the decay constant for deuterium beta decay. The $pn$ reaction rate is taken from \cite{Smith1993,Arbey2012}.

We solve the equilibrium conditions above for the number densities of free neutrons and deuterium. By defining the following dimensionless parameters,
\begin{align}
f_1 &= \frac{n_p \sv_{pn}}{n_p \sv_{Dp} + \lambda_D + \lambda^\beta_D} \\
f_2 &= \frac{\frac{1}{2} n_p \sv_{pp} +  n_p n_e \Kp}{n_p \sv_{Dp} + \lambda_D + \lambda^\beta_D} \\
f_3 &= \frac{\lambda_D}{n_p \sv_{pn} + \lambda_n} ~,
\end{align}
the equilibrium abundances are,
\begin{equation}
\frac{n_D}{n_p} = \frac{f_2}{1 - f_1 f_3} ~ , \qquad \frac{n_n}{n_p} = \frac{f_2 f_3}{1 - f_1 f_3}
\end{equation}
which simplifies to the abundance in stars like the Sun ($n_D/n_p = \sv_{pp} ~/~ 2\sv_{Dp}$) if the deuteron is stable ($\lambda_D = \lambda^\beta_D = 0$) and the $pep$ reaction is negligible ($\Kp = 0$). The rate of \HeT \emph{production} is,
\begin{equation} \label{eq:dHe3dt}
\left. \frac{\df n_{\HeT}}{\df t} \right\rvert_\ro{prod} = n_p n_D \sv_{Dp},
\end{equation}
and \HeT nuclei are quickly converted into \HeF via Reaction \eqref{eq:React4},
\begin{equation} \label{eq:dHe4dt}
\frac{\df n_{\HeF}}{\df t} = \frac{1}{2}  \left. \frac{\df n_{\HeT}}{\df t} \right\rvert_\ro{prod}  ~.
\end{equation}
The energy released by the net reaction $4p + 2 e^- \rightarrow \HeF + 2 \nu_e$ (recalling that the positrons created in Reaction \eqref{eq:React1} will quickly annihilate with electrons), is $Q_{\HeF} = 4 m_p + 2m_e - m_{\HeF} = B_{\HeF} + 2m_e - 2(m_n - m_p)$. In our universe, $Q_{\HeF} = 26.73$ MeV.  We will assume that the triple alpha rate $3 \HeF \rightarrow \CTw$ is negligible during the proton-burning phase.

We can write the nuclear energy production rate (energy per unit time, per unit volume) from the $4p$ reaction as,
\begin{equation} \label{eq:eps4p}
\epsilon_{4p} = Q_{\HeF} \frac{\df n_{\HeF}}{\df t} ~.
\end{equation}
We will keep $Q_{\HeF}$ constant throughout this work; our changes in binding energy for the deuteron are small compared to the binding energy of \HeF.

\subsection{General \emph{pp}-Reaction Rate}
The reaction rate per particle pair for an \emph{exothermic} two-body reaction $1 + 2 \rightarrow 3 + 4$ can be written in terms of astrophysical $S$-factors as,
\begin{equation} \label{eq:sv}
\sv_{12} = \frac{8}{\sqrt{3} \pi Z_1 Z_2 \alpha \mR c} S_0 \Theta^2 \exp (-3 \Theta) ~.
\end{equation}
However, when the $pp$ reaction is endothermic, we need to revisit the assumptions that lead to Equation \eqref{eq:sv}. In the $pp$ reaction, the final kinetic energy of the electron and neutrino produced in the $pp$ reaction is $K_f = K_i + Q_{pp}$, where $K_i$ is the initial centre-of-mass frame kinetic energy of the colliding protons. In our universe, $Q_{pp} =  0.420$ MeV, and at the temperature of the core of the sun ($2 \ten{7}$ K), $K_i \approx k T_\ro{core} = 0.002$ MeV. Thus, the initial kinetic energy of the protons can be ignored, and $S_{pp}$ is approximately independent of temperature.

As $Q_{pp}$ is reduced, the initial kinetic energy of the protons becomes more important. And if $Q_{pp}$ is negative, then there is a kinetic energy threshold: the reaction will only occur if $K_i > -Q_{pp}$. This produces an exponential suppression of the reaction at small temperatures, as only rare protons in the high-energy tail of the Maxwell-Boltzmann distribution are able to produce deuterons.

With this understanding of the importance of the initial kinetic energy of the protons, we can use the standard calculation of $S_{pp}$ from the theory of the weak nuclear force \cite{Salpeter1952,Iben2013},
\begin{equation} \label{eq:Spp}
S_{pp}(K_i) = \frac{3}{4 \pi^2 }\frac{\alpha c^5 m_p m_e^5}{\hbar^7} g^2_{weak} s^2 I_d^2 F_{pp}(K_i) ~,
\end{equation}
where $K_i$ is the centre-of-mass initial kinetic energy of the protons, $g_{weak} = 1.25 G_F = 1.80 \ten{-49} $ erg cm$^3$, $s \approx 1.93$ and $I_d$ are parameters relating to the nuclear overlap integral. The calculation of $s$ requires $V_0$, the depth of the nuclear potential well. It is known that $V_0 \sim 20$ MeV; we set $V_0 \sim 18.1$ MeV, so that the resulting value of $S_{pp}$ agrees with the observed value. Since we are only changing the binding energy of deuterium by an amount of order its value in our universe (2.2 MeV), we assume that the overall depth of the well stays approximately constant. Only small changes are needed to alter the energy of the bound $D$ state.

The factor $F_{pp}$, analogously to the Fermi function, is a dimensionless function of the energy of the products of the reaction. This is where the dependence on $K_i$ enters the equation,
\begin{equation}
F_{pp}(K_i) = -\frac{1}{4} \eta_f - \frac{1}{12} \eta_f^3 + \frac{1}{30} \eta_f^5 + \frac{1}{4} \epsilon_f ~ \log_e(\eta_f + \epsilon_f )
\end{equation}
where $\epsilon_f =  (K_i + Q_{pp} + m_e)/m_e$ and $\eta_f = (\epsilon_f^2 - 1)^{1/2}$. For most stars in our universe,  $K_i \ll Q_{pp}$, which makes the S-factor an approximately constant function of energy. 

To calculate the reaction rate of the $pp$ reaction at a given temperature, we must average over the distribution of initial kinetic energies. From \cite{Iben2013}, Equation (6.6.10),
\begin{equation} \label{eq:Rpp}
R_{pp} = \frac{n_p^2}{2} \frac{4}{(2 \pi \mR kT)^{1/2}} 
			\int_{x_{min}}^\infty S_{pp}(K_i) \exp \left(-x - \frac{\lambda}{x^{1/2}} \right) \df x
\end{equation}
where $x = K_i/kT$, $\mR = 1/2m_p$ is the reduced mass of the reactants, $\lambda = 2 \pi \alpha (\mR c^2/2kT)^{1/2}$ is related to the Gamow energy, and $x_{min} = \ro{max}(-Q_{pp}/kT,0)$ takes into account the kinetic energy threshold, effectively setting $S_{pp}$ equal to zero if $K_i <  -Q_{pp}$.  Because we cannot consider $S_{pp}$ to be a constant, and because of the reaction threshold for an endothermic reaction, we cannot approximate the integral by a Gaussian, as is usually done to give the expression in Equation \eqref{eq:sv}. Instead, the integral is evaluated numerically for each change in fundamental parameters.

Figure \ref{fig:RppKpep} (left) shows the ratio of the reaction rate (per $n_p^2$) at a given value of $Q_{pp}$ relative to its value in our universe. For positive values of $Q_{pp}$, the rate is reduced, but still asymptotes to a constant value at low temperature where the initial kinetic energy of the reactants is negligible. For negative values of $Q_{pp}$, the reaction rate is exponentially suppressed at low temperature.

\subsection{General \emph{pep}-Reaction Rate}
For the three-body $pep$ reaction, we use the formalism of \cite{Irgaziev2014}, where the rate constant \Kp for the $pep$ reaction as a function of temperature (T) is given by,
\begin{equation} \label{eq:kpep}
\Kp = \frac{1}{(kT)^3} \int_{K_\ro{min}}^\infty G_0(K_i) ~ S_{pep}(K_i) ~ \ro{e}^{-K_i/kT} ~K_i^2 \df K_i
\end{equation}
where $K_i$ is the total kinetic energy of the reactants (denoted $E$ in \cite{Irgaziev2014}), $K_\ro{min} = \ro{max}(-Q_{pep},0)$ takes into account the energy threshold if the reaction is endothermic,  $G_0$ is the Gamow factor for the $pep$ reaction, and $S_{pep}$ is the astrophysical S-factor, which is a shallow function of energy. By Equation (26) of \cite{Irgaziev2014}, $S_{pep}$ is proportional to $E_\nu^2 = (K_i + Q_{pep})^2$. Effectively, the overlap integral of the wavefunctions is unaffected by the changes to the parameters we consider, so the only change results from the density of states. We numerically calculate the integral in Equation \ref{eq:kpep}.

\begin{figure*} \centering
	\begin{minipage}{0.48\textwidth}
		\includegraphics[width=\textwidth]{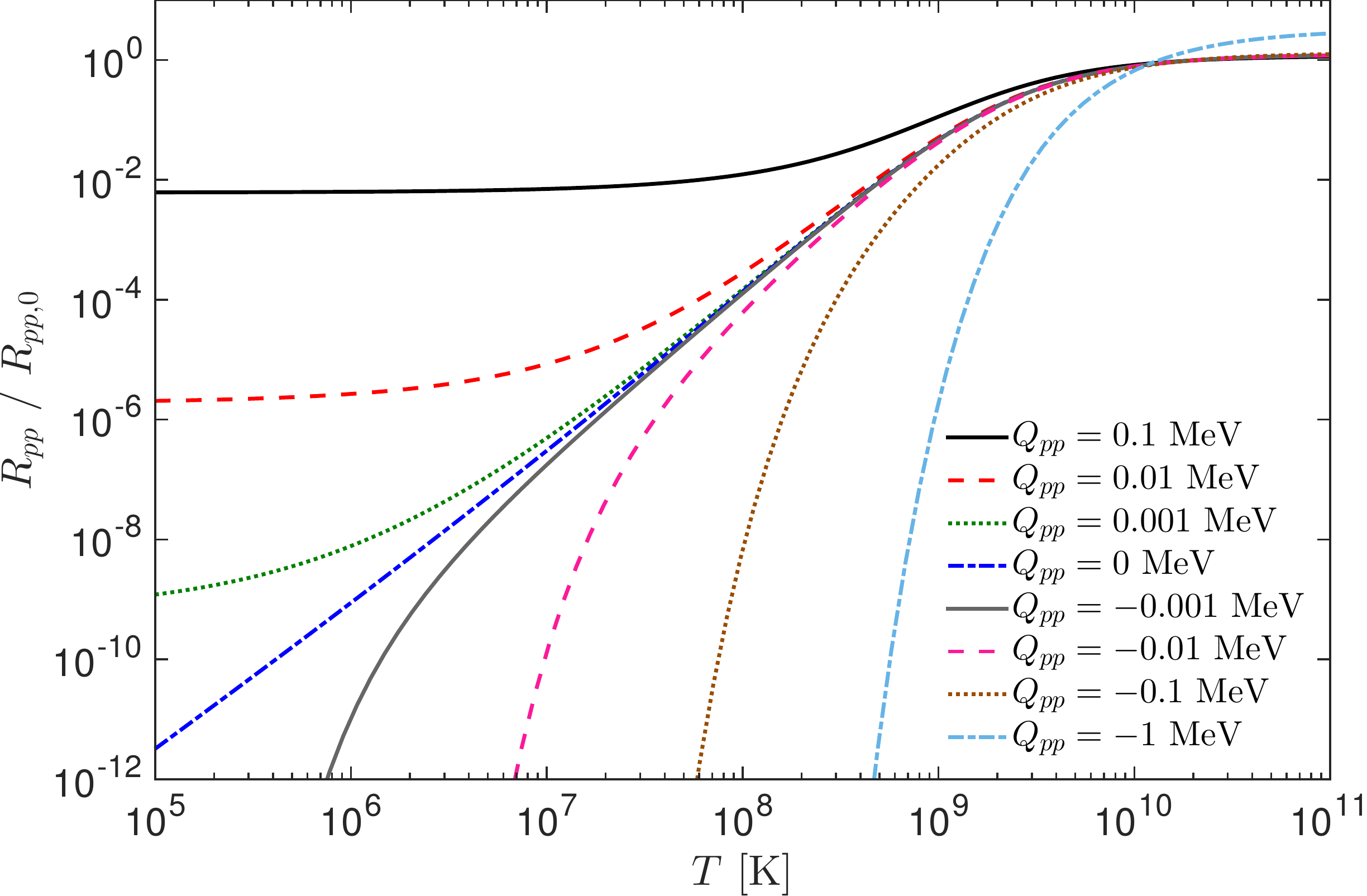}
	\end{minipage}
	\begin{minipage}{0.48\textwidth}
		\includegraphics[width=\textwidth]{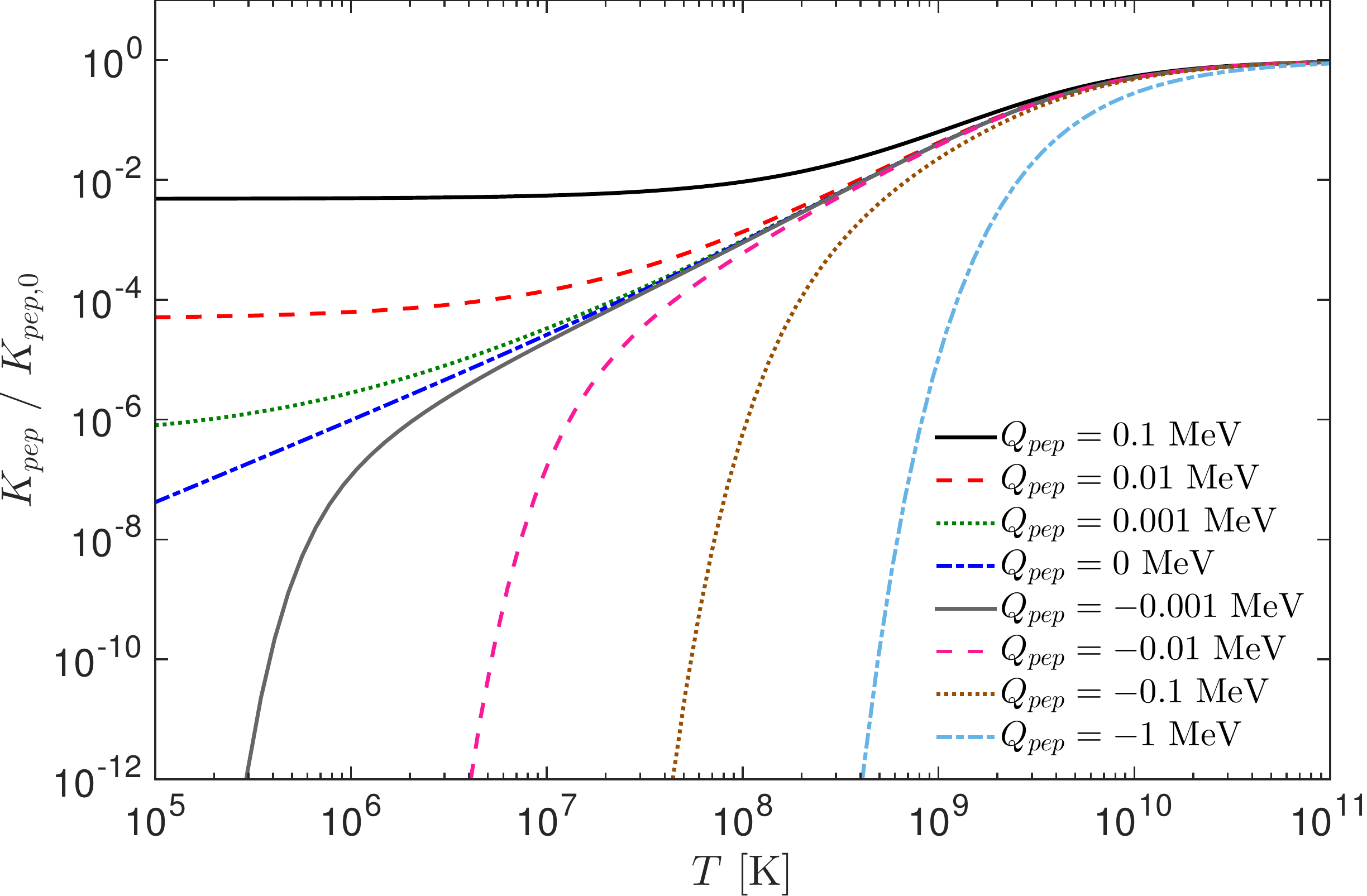}
	\end{minipage}
\caption{\emph{Left:} the ratio of the $pp$ reaction rate (per $n_p^2$) at a given value of $Q_{pp}$ relative to its value in our universe, as a function of temperature. For negative values of $Q_{pp}$, a kinetic energy threshold for the reactants results in a low-temperature exponential suppression. \emph{Right:} the ratio of the $pep$ reaction rate (per $n_p^2 n_e$) at a given value of $Q_{pep}$ relative to its value in our universe, as a function of temperature. For negative values of $Q_{pep}$, a kinetic energy threshold for the reactants results in a low-temperature exponential suppression.}
\label{fig:RppKpep}
\end{figure*}

Figure \ref{fig:RppKpep} (right) shows the ratio of the reaction rate (per $n_p^2 n_e$) at a given value of $Q_{pep}$, relative to its value in our universe. As with the $pp$ reaction rate, positive values of $Q_{pep}$, the rate is reduced, but still asymptotes to a constant value at low temperature. For negative values of $Q_{pep}$, the reaction rate is exponentially suppressed at low temperature due to the kinetic energy threshold.

\subsection{Neutrino Losses}
In calculating the net energy production of our reaction pathway, there is an important complication. The reaction pathway we are considering can produce a neutrino/antineutrino pair even if it doesn't produce a  \HeF nucleus. This pair will not annihilate but instead escape the star, carrying away energy. In proton-burning stars in our universe, neutrino production slightly lowers the effective energy released by each reaction. But if stars produce heat energy in a three-body process, but lose energy to neutrinos as the results of a two body process, then the situation is more complicated.

Taking neutrino losses into account, the total rate at which heat energy is injected into the star from its nuclear reserve is,
\begin{equation} \label{eq:epstot}
\epsilon_{\ro{net}} = \epsilon_{4p} - \epsilon^\nu_{pp} - \epsilon^\nu_{pep} - \epsilon^\nu_{D} - \epsilon^\nu_{n} ~,
\end{equation}
where the negative terms are the energy lost to the neutrino produced in the $pp$ reaction, $pep$ reaction, $D$ beta decay and neutron beta decay respectively, as a function of temperature.

Where the initial kinetic energy of the reactants is negligible, the $pp$ and $pep$ terms are given by a typical energy per reaction ($Q^\nu_{pp}$ and $Q^\nu_{pep}$) multiplied by the reaction rate. Note that if the deuteron is stable, the $pp$ and $pep$ neutrino loss terms simply reduce the overall nuclear $Q$ value of the corresponding reactions. In the presence of an unstable deuteron, however, neutrino losses can be substantial. For example, in the extreme case of an very unstable deuteron, the $pp$, $pep$ and $D$-decay reactions will simply cycle, producing little helium and radiating copious energy in the form of neutrinos. Very high temperatures may be required for the first term above to dominate, so that $\epsilon_{\ro{net}} > 0$. Note also that, even in this case, $\epsilon_{\ro{net}}$ could change sign in the less dense, cooler outskirts of the star.

We are considering cases where the initial kinetic energy of the reactants ($K_i$) is not negligible, so we need to calculate these terms in full generality. We calculate the average energy of the neutrino produced, as a function of $K_i$ as follows. It is easiest for the $pep$ reaction, since the neutrino carries away almost all of the energy from the reaction: $Q^\nu_{pep} = K_i+Q_{pep}$.

For the $pp$ reaction, the final kinetic energy is shared between the electron and the neutrino. We can calculate the spectrum of energy of the electron and neutrino using beta decay theory (e.g. \cite{Bertulani2007}, Chapter 8). The electron kinetic energy distribution is given by,
\begin{equation}
p(E_e) dE_e \propto F(Z,E_e) ~E_\nu^2 ~p_e ~ (E_e + m_e c^2)  \df E_e ~,
\end{equation}
where $F(Z,E_e)$ is the Fermi Function \cite{NBS1952}, $Z$ is the atomic number of the decay product, $E_\nu$ is the energy of the neutrino and $p_e$ is the momentum of the electron. The maximum value of $E_e$ and $E_\nu$, by conservation of energy, is $E^{pp}_{max} = K_i + Q_{pp}$. Then,
\begin{align}
Q^\nu_{pp} =& \langle E_\nu \rangle \\
=& E^{pp}_{max} - \langle E_e \rangle \\
=& E^{pp}_{max} - \int_0^{E^{pp}_{max}} E_e ~ p(E_e) \df E_e ~.
\end{align}
We consider the $pp$ reaction to be a diproton beta-decay, as far as the spectrum of neutrinos is concerned. Setting $Z = 1$, and given that $K_i$ is negligible in our universe, we correctly calculate the experimental result $\langle E_\nu \rangle = 0.2668$ MeV \cite{Bahcall2002}.

To calculate the neutrino losses as a function of temperature,  we need to average the quantities $Q^\nu_{pp}$ and $Q^\nu_{pep}$ over the distribution of interaction energy\footnote{To be precise, we should do the same for the energy released from the $pp$ reaction, and its contribution to $Q_{\HeF}$ in Equation \eqref{eq:eps4p}. We can ignore this effect as the dominant contribution to $Q_{\HeF}$ is from the $D + p$ and $\HeT + \HeT$ reactions, which are unaffected by the changes to the constants that we consider here.}. We do this in a similar way to calculating the $pp$ (Equation \ref{eq:Rpp}) and $pep$ (Equation \ref{eq:kpep}) reaction rates at a given temperature, averaging with respect to the reaction rate, replacing the corresponding terms in $\epsilon_{\ro{net}}$ \eqref{eq:epstot},
\begin{align}
\epsilon^\nu_{pp} &= \frac{n_p^2}{2} \frac{4}{(2 \pi \mR kT)^{1/2}} \int_{x_{min}}^\infty Q^\nu_{pp}(K_i) S_{pp}(K_i) \exp \left(-x - \frac{\lambda}{x^{1/2}} \right) \df x ~, \\
\epsilon^\nu_{pep} &= \frac{n_p^2 n_e}{(kT)^3} \int_{K_\ro{min}}^\infty Q^\nu_{pep}(K_i)~ G_0(K_i) ~ S_{pep}(K_i) ~ \ro{e}^{-K_i/kT} ~K_i^2 \df K_i ~.
\end{align}
For regimes in $Q^\nu_{pp}(K_i)$ and $Q^\nu_{pep}(K_i)$ that are approximately constant functions of the initial kinetic energy, these expressions simplify to the energy per reaction multiplied by the reaction rate: $\epsilon^\nu_{pp} = Q^\nu_{pp} \frac{1}{2} n_p^2 \sv_{pp}$, $\epsilon^\nu_{pep} = Q^\nu_{pep} n_p^2 n_e \Kp$.

For the decay of a free neutron, $\epsilon^\nu_{n} = Q^\nu_{n} n_n \lambda_n$, where $Q^\nu_{n} = 0.5$ MeV is the average energy of the emitted neutrino. We can calculate the decay constant $\lambda^\beta_D$ and the average energy carried away by the neutrino in $D$ beta decay by applying standard beta decay theory to Reaction \eqref{eq:React2b}. The calculation of $Q^\nu_{D}$ is then similar to the calculation of $Q^\nu_{pp}$, setting $Z = 1$ and the overlap integral to unity. The rate of neutrino loss is then $\epsilon^\nu_{D} =  Q^\nu_{D} n_D \lambda^\beta_D$. We tested our code by calculating the lifetime of tritium, which has a lifetime of 12.32 years, long enough to be measured in underground water sources \cite{JB1971,JB1973}. 

\subsection{Solving for the Central Temperature}
Following Adams \cite{Adams2008}, we complete our system of equations by considering the energy transport equation,
\begin{equation} \label{eq:T3dr}
T^3 \frac{\dd T}{\dd r} = -\frac{3 \rho \kappa}{4 a c} \frac{L(r)}{4 \pi r^2} ~,
\end{equation}
where $\kappa$ is the opacity, and $L(r)$ is the net energy passing through a sphere of radius $r$ per unit time, which in equilibrium will be equal to the net energy produced inside $r$ per unit time,
\begin{equation} \label{eq:Lr}
L(r) = \int_0^r \epsilon_{\ro{net}} (r')  4 \pi r'^2 \df r'
\end{equation}
Following Adams, we make the simplifying assumption that the combination $\kappa \rho = \kappa_0 \rho_c$ is constant, which gives,
\begin{equation}
\int_0^{R_*} \frac{L(r)}{4 \pi r^2} \df r =  \frac{a c T_c^4}{3 \rho_c \kappa_0} ~.
\end{equation}
This is, in fact, an implicit equation that we can solve to find the central temperature $T_c$ for a star of a given mass $M_*$. Unlike Adams, we do not attempt to simplify this equation by assuming a central power-law reaction rate. We calculate the integrals numerically\footnote{We can simplify the double integral using integration by parts: 
\begin{equation}
\int_0^{R} \frac{1}{r^2} \left[ \int_0^{r} f (r')  r'^2 \df r' \right] \df r = \int_0^{R} \left( r - \frac{r^2}{R}\right) f(r) \df r ~.
\end{equation}}.

Having found $T_c$, we can calculate the stellar radius $R_*$, luminosity $L_*$ (via Equation \ref{eq:Lr}), surface temperature $T_*$, and an approximate upper limit for the main-sequence lifetime of a star $t_*$,
\begin{align}
R_* &= \frac{G M_* \bar{m}} {k_B T_c} \frac{\xi_*} {(n+1) \mu_0} \\
T_* &= \left( \frac{L_*} {4 \pi R_*^2 \sigma_{\!_{SB}}} \right)^{1/4} \\
t_* &= \frac{\epsilon_{\ro{nuc}} M_* c^2} {L_* + L_\nu} ~.
\end{align}
where $\sigma_{\!_{SB}}$ is the Stefan-Boltzmann constant and $\epsilon_{\ro{nuc}}$ is the ratio of the stars available nuclear energy to the rest mass. Note that the calculation of the lifetime depends on the rate at which all forms of energy are produced (light and neutrinos), but doesn't take into account the effects of stellar evolution, and assumes that the star burns all of the available nuclear fuel.

The free parameters of our model are listed in Table \ref{tab:table_params}, along with their fiducial values. For simplicity, we consider pure hydrogen stars ($Y_\ro{He} = 0$), and assume that the star is ionized throughout. Using the equations above, we can calculate the lower and upper mass limits of stable stars. As in B15, the lower limit is derived from the onset of electron degeneracy pressure, such that small balls of gas are stabilized at temperatures too low for nuclear reactions to ignite. The upper limit comes from the onset of radiation pressure domination, which results in very short lived, marginally stable stars. The formulae for these limits can be found in B15; they are unchanged by our alterations to nuclear physics. However, the possibility of large neutrino losses means that stars may also fail to be stable because $L(r)$ becomes negative at large radii, which is not permitted by Equation \ref{eq:T3dr}. For some models considered below, this provides a stronger lower limit on the set of possible stellar masses.

\begin{table*}
\centering
\begin{tabular}{|c|c|c|}
\hline
Model Parameter & Symbol							 					& In our Universe  \\ 
\hline
Polytropic index    							& $n$ 					  	& 3/2   \\
Gamow energy, $pp$ reaction      	& $E_{pp}$	  			& 0.493 MeV   \\
Gamow energy, $Dp$ reaction      	& $E_{Dp}$ 	  			& 0.658 MeV   \\
$pep$ reaction energy      				& $Q_{pep}$ 			& 1.442 MeV   \\
$pp$ reaction energy       				& $Q_{pp}$ 				& 0.420 MeV   \\
$4p\rightarrow \HeF$ reaction energy & $Q_{\HeF}$ 		& 26.73 MeV   \\
Burning efficiency  							& $\epsilon_\ro{nuc}$ & 0.0071 \\
$pp$ neutrino energy loss 				& $Q^\nu_{pp}$ 		& 0.2668 MeV  \cite{Bahcall2002} \\
$pp$ neutrino energy loss 				& $Q^\nu_{D}$	 		& 0   \\
Deuteron binding energy 				& $B_D$					& 2.2245 MeV   \\
Deuteron unbinding decay constant	& $\lambda_D$  	& 0   \\
Deuteron beta decay constant	 		& $\lambda^\beta_D$	& 0   \\
Hydrogen mass fraction & $X_H$							      	& 1   \\
Helium mass fraction & $Y_\ro{He}$							  	& negligible   \\
Initial deuterium number fraction &  \Deu  					  	& negligible   \\
$pp$ astrophysical S-factor &	 $S_{pp}$						 	& $4.01 \ten{-25}$ MeV b  \cite{Adelberger2011}   \\
$Dp$ astrophysical S-factor &	 $S_{Dp}$						 	& $2.14 \ten{-7}$  MeV b   \cite{Adelberger2011}   \\
Central opacity	 		& $\kappa_0$									& 0.398 cm$^2$ g$^{-1}$  \\
Critical gas/total pressure & $f_g$								  	& 0.5   \\
Average particle mass 		& $\bar{m}$ 						  	& 0.5 \mPr   \\
Mass per free electron     & $\mion$					  	  	  	& \mPr   \\
\hline 
\end{tabular}
\caption{Free parameters of our stellar model, and our choices of parameters for our fiducial unstable-D star. The polytropic index of stars in our universe is generally believed to increase from $n = 3/2$ for small stars to $n =3$ for high mass stars; the effect on our modelling is small, so we will set $n = 3/2$. The central opacity  is approximated using the Thomson cross-section: $\kappa_0 = \sigma_T / \mion$.}
\label{tab:table_params}
\end{table*}

\section{Unstable Deuteron Stars} \label{S:UnstableD}

To illustrate the effect of an unstable Deuteron on stars, we consider a simplified case in which we take the fiducial model of Table \ref{tab:table_params} and add a non-zero value for $\lambda_D$. In this section we will ignore the $pep$ reaction by setting $\Kp = 0$. Note that many of the parameters in our model depend on the masses of the light quarks and electron, as well as the strengths of the fundamental forces, so varying only the Deuteron decay constant is somewhat artificial. It will, regardless, illustrate the relevant effects. We will also illustrate the effects of neutrino losses by turning them completely on or off.

\begin{figure*} \centering
	\begin{minipage}{0.45\textwidth}
		\includegraphics[width=\textwidth]{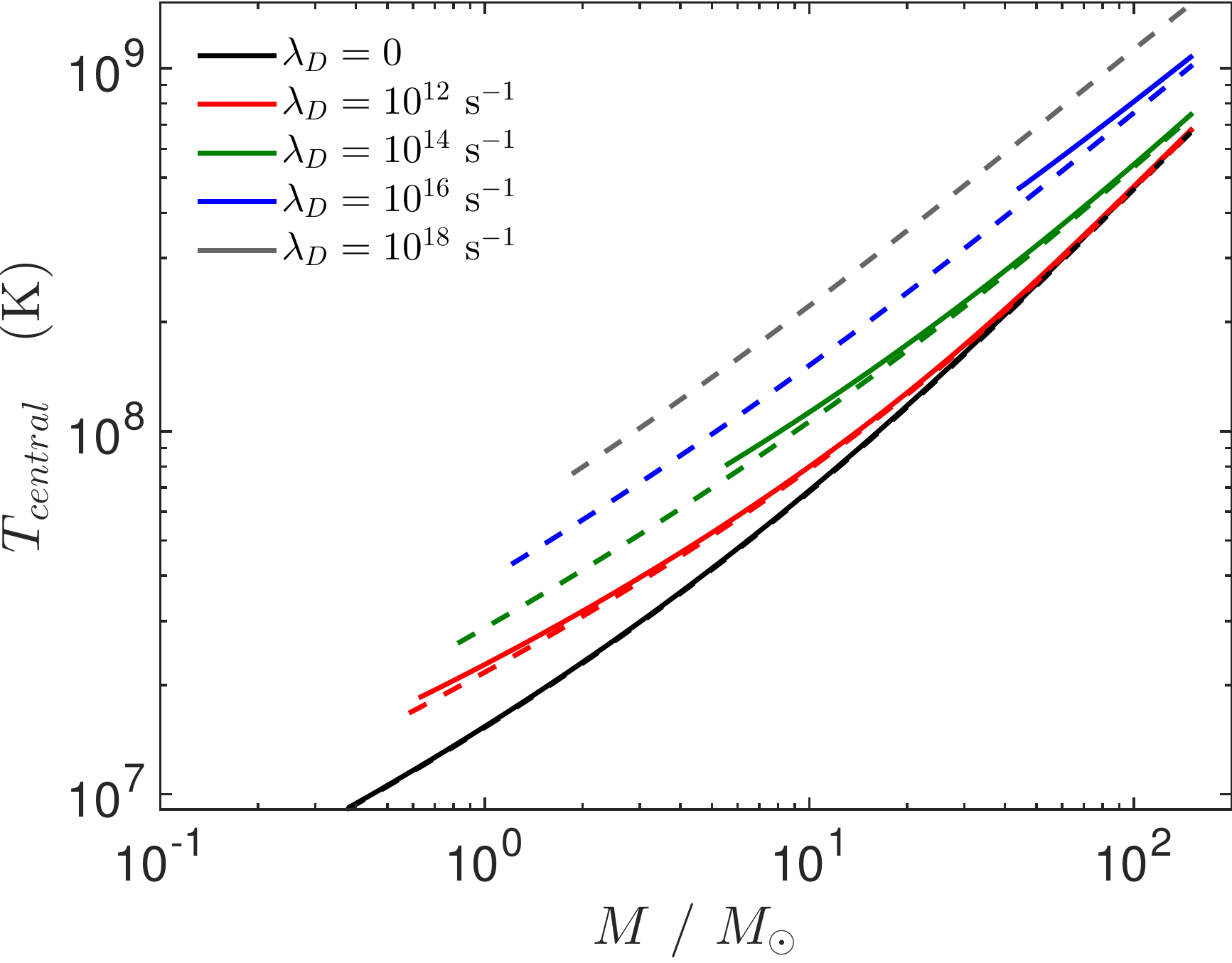}
	\end{minipage}
	\begin{minipage}{0.45\textwidth}
		\includegraphics[width=\textwidth]{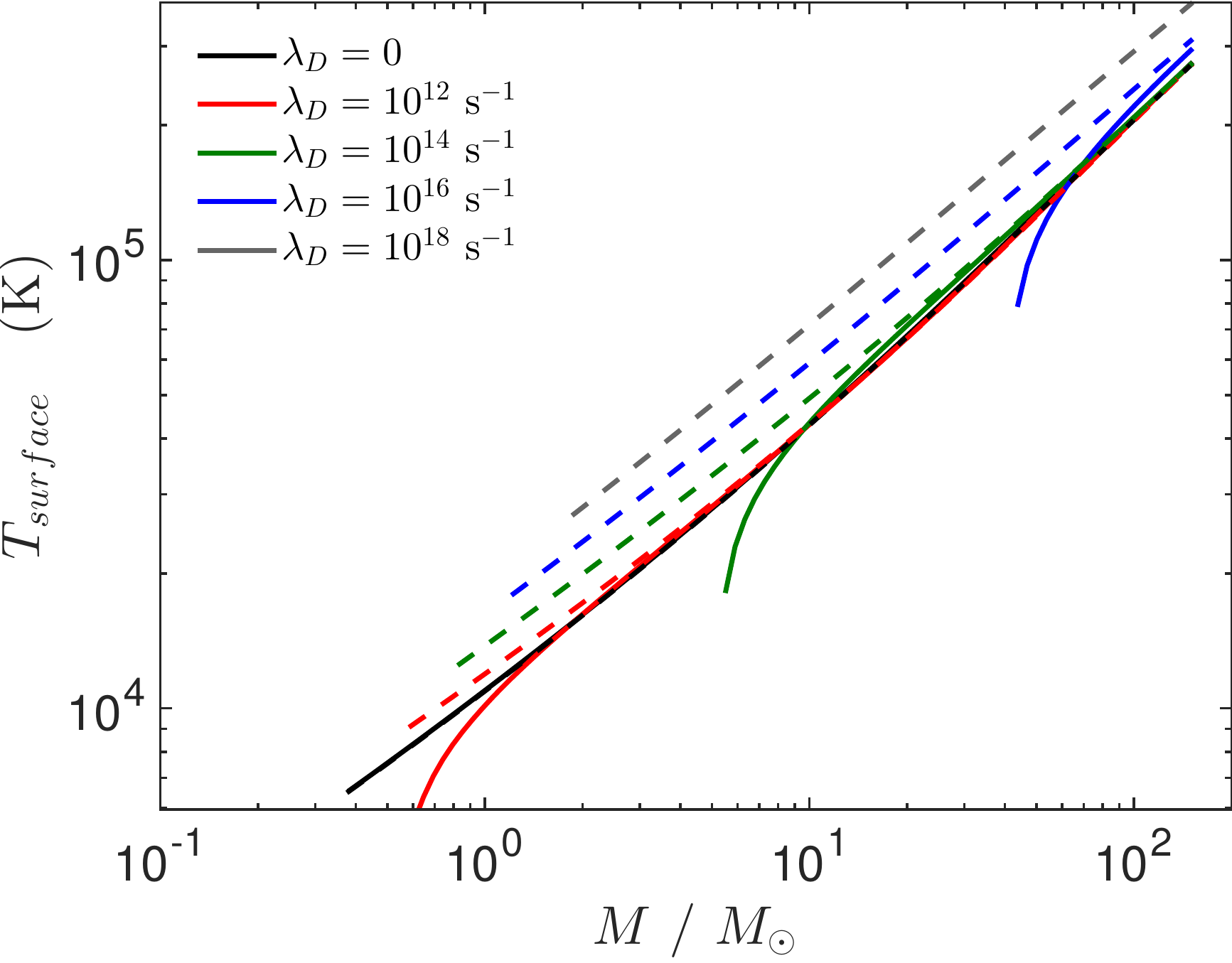}
	\end{minipage}
	\begin{minipage}{0.45\textwidth}
		\includegraphics[width=\textwidth]{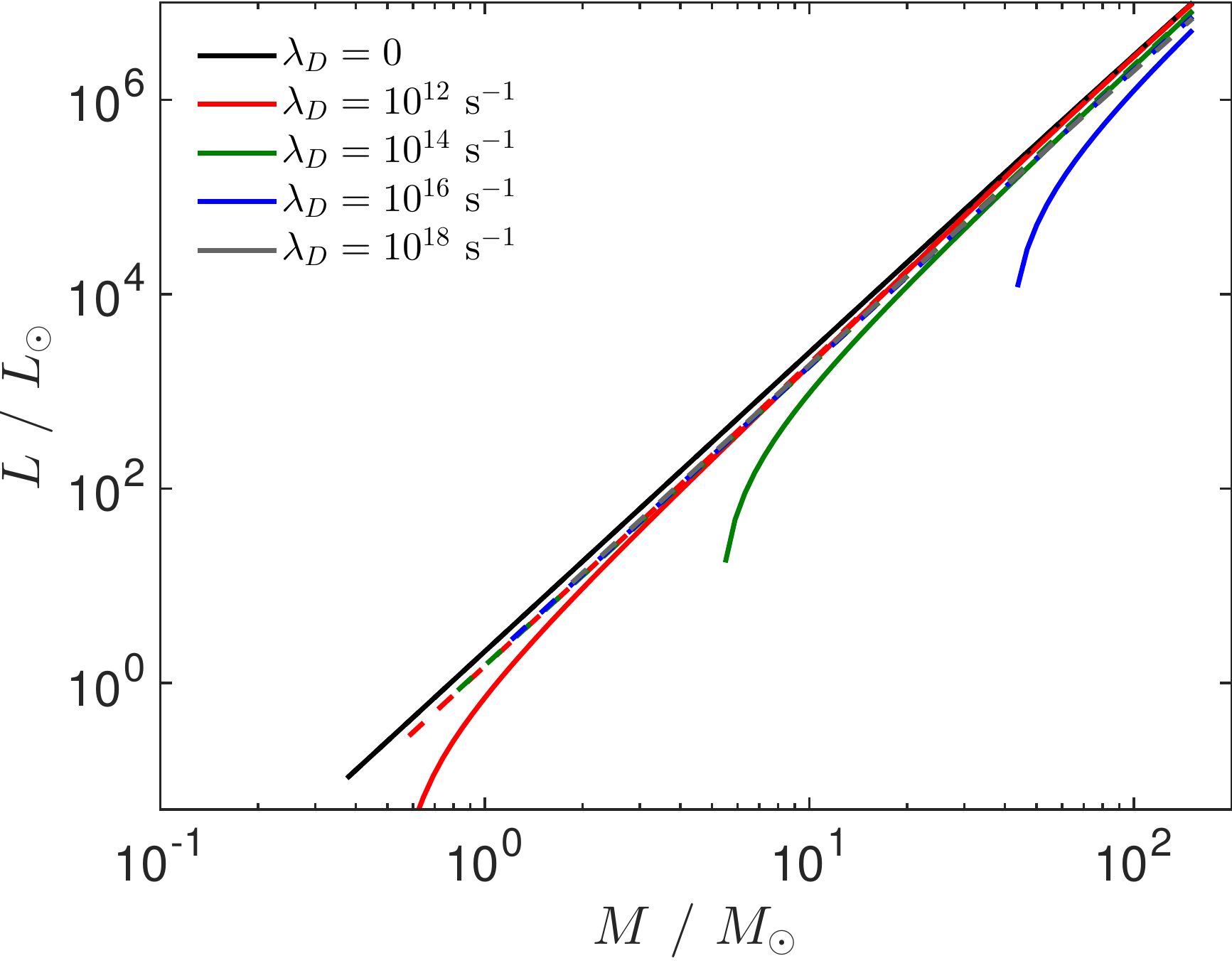}
	\end{minipage}
	\begin{minipage}{0.45\textwidth}
		\includegraphics[width=\textwidth]{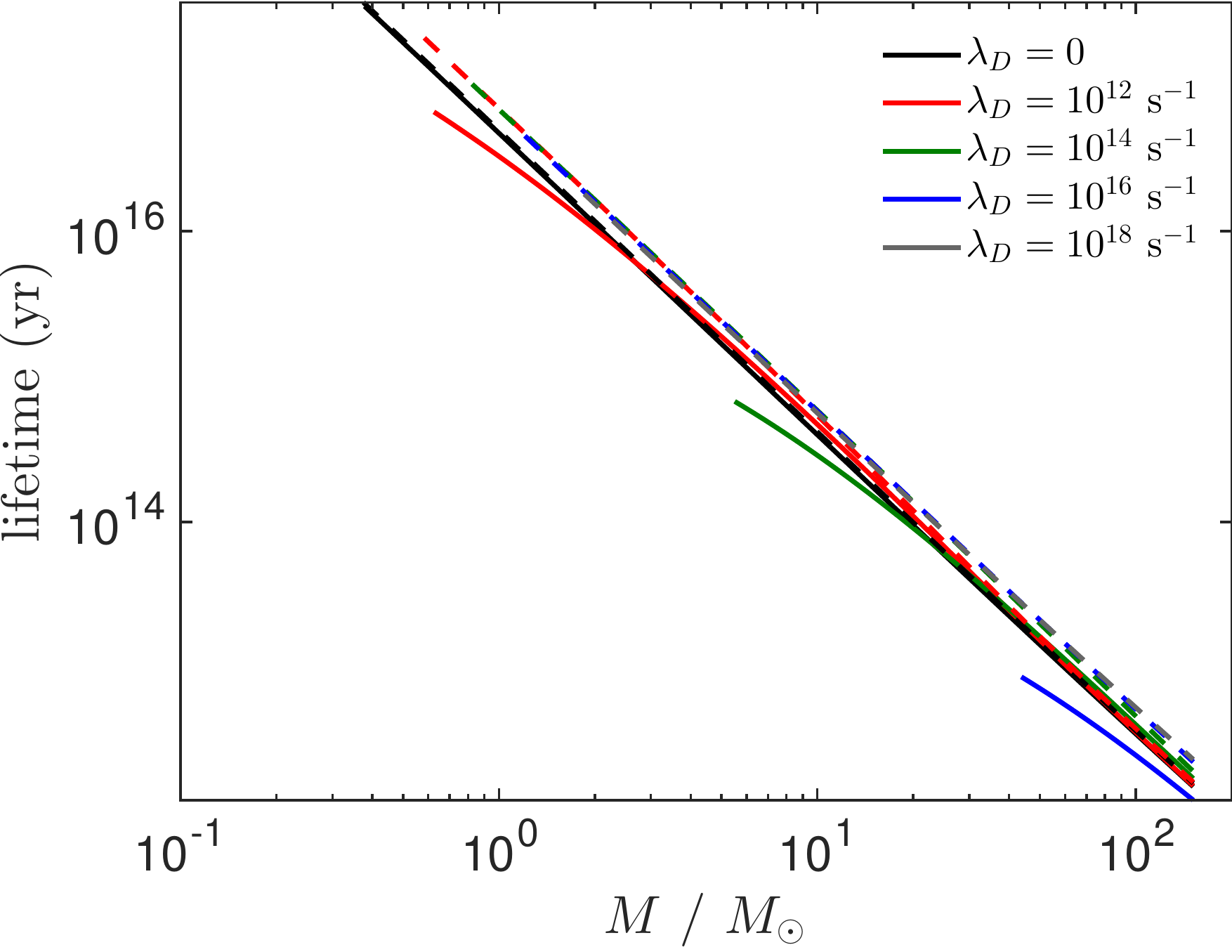}
	\end{minipage}
\caption{Stars in universes in which the lifetime of the deuteron is increasingly short. The different colours of line show different values of the deuterium decay constant: $\lambda_D = (0,10^{12},10^{14},10^{16},10^{18})$ s$^{-1}$, from the lowermost to uppermost line respectively. The dashed lines show the effect of neglecting neutrino losses. Note that the solid grey line ($\lambda_D = 10^{18}$ s$^{-1}$) is absent --- when neutrino losses are taken into account, stars are not possible with such a small value of the deuteron half life.}
\label{fig:Dmodels}
\end{figure*}

Figure \ref{fig:Dmodels} shows the effect of different values of the deuterium decay constant: $\lambda_D = (0,10^{12},10^{14},10^{16},10^{18})$ s$^{-1}$, from the lowermost (black) to uppermost (grey) solid line respectively. The dashed lines show the effect of neglecting neutrino losses. Note that the solid grey line ($\lambda_D = 10^{18}$ s$^{-1}$) is absent --- when neutrino losses are taken into account, stars are not possible with such a small value of the deuteron half life.

The upper left panel shows the relationship between central temperature and stellar mass. As expected, at a fixed stellar mass, as the half-life of the deuteron (ln2 $/\lambda_D$) decreases, the star must burn hotter at its centre to maintain equilibrium. But perhaps the most obvious effect of this increased ignition temperature is that smaller stars are unable to ignite, and the minimum stellar mass increases as the deuteron becomes more stable. The upper mass limit is not similarly affected, and so the stellar window closes as $\lambda_D$ increases.

The surface temperature (upper right) and stellar (radiation) luminosity (lower left) show the effect of neutrino losses in absorbing energy in the outer parts of the star. Apart from these declines at low masses, an unstable deuteron has minimal effect on the luminosity of the star at fixed stellar mass. The reason is that the luminosity of the star is determined not primarily by the nuclear reactions, but by the required balance between thermal pressure and gravity, and the radiation transport of thermal energy. Similarly to the stable diproton stars of B15, the central density and temperature (and hence reaction rate) adjusts to maintain overall thermodynamic equilibrium.

The lifetime of the star (bottom right) is also lowered by neutrino losses, with the corresponding curves dipping below the overall, deuterium half-life independent trend. Recall that the lifetime depends on the total luminosity, including neutrinos, so the lower radiation luminosity (bottom left) does not imply a longer lifetime.

The increasing minimum mass with increasing instability means that the maximum stellar lifetime decreases with increasing $\lambda_D$. At the extremes of the range of $\lambda_D$ that we consider, only very large stars are able to ignite, and so the longest lived stars burn out in millions of years. For $\lambda_D \gtrsim 10^{17}$ s$^{-1}$ and taking neutrino losses into account, the stellar window closes completely --- stable, self-gravitating balls of gas are unable to ignite nuclear reactions. Note that this timescale $10^{17}$ s$^{-1}$ is similar to the lifetime of the unstable $\ce{^{8}Be}$ nucleus. Note, however, that $\ce{^{8}Be}$ is bound, with a binding energy of $\sim 7$ MeV per nucleon. It decays because the substantial binding energy of \HeF makes the alpha-decay $\ce{^{8}Be} \rightarrow \HeF + \HeF$ energetically favourable. An unbound deuteron, as noted above, does not decay by quantum tunneling through a barrier, and so its lifetime is approximately equal to the time taken for the nucleons to travel across the nuclear potential well ($\sim 10^{-23}$ seconds). This is similar to the lifetimes of $\ce{^{4}Li}$ and $\ce{^{5}Li}$, which decay by simply ejecting a proton.

The dashed lines show the effect of neglecting neutrino losses. As expected, at fixed stellar mass, the presence of neutrino losses demands a higher central temperature. Moreover, the effect of neutrino losses is more pronounced for larger values of $\lambda_D$, as a result of the larger contribution of neutrinos from deuteron decay.

On quite general grounds, then, we can conclude that a universe with an unbound deuteron, with $\lambda_D \sim 10^{23}$ s$^{-1}$, will not contain nuclear-burning stars.

\section{A Less Bound Deuteron} \label{S:endopp}
We will investigate the effects of a less-bound deuteron by changing $B_D$ only. This can be achieved by changing the strength of the strong force, leaving the other fundamental constants of nature unaltered. These universes are below ours in Figure \ref{fig:Dme}. 

To illustrate the effect of decreasing the binding energy of the deuteron, and the resulting endothermic $pp$ and $pep$ reactions, we consider the following values of $\Delta E = B_D - (m_n - m_p) = (0.931,0.6,0.2,-0.2,-0.6,-1.0,-1.283)$. The first is the value of $\Delta E$ in our universe, and the final is for a marginally bound deuteron $B_D = 0.01$ MeV. The first two values are in the upper region of Figure \ref{fig:Dme} ($pp$ and $pep$ exothermic), the second two are in the middle region ($pp$ endothermic, $pep$ exothermic), and the final three values are in the lower region ($pp$ and $pep$ endothermic).

\begin{figure*} \centering
	\begin{minipage}{0.45\textwidth}
		\includegraphics[width=\textwidth]{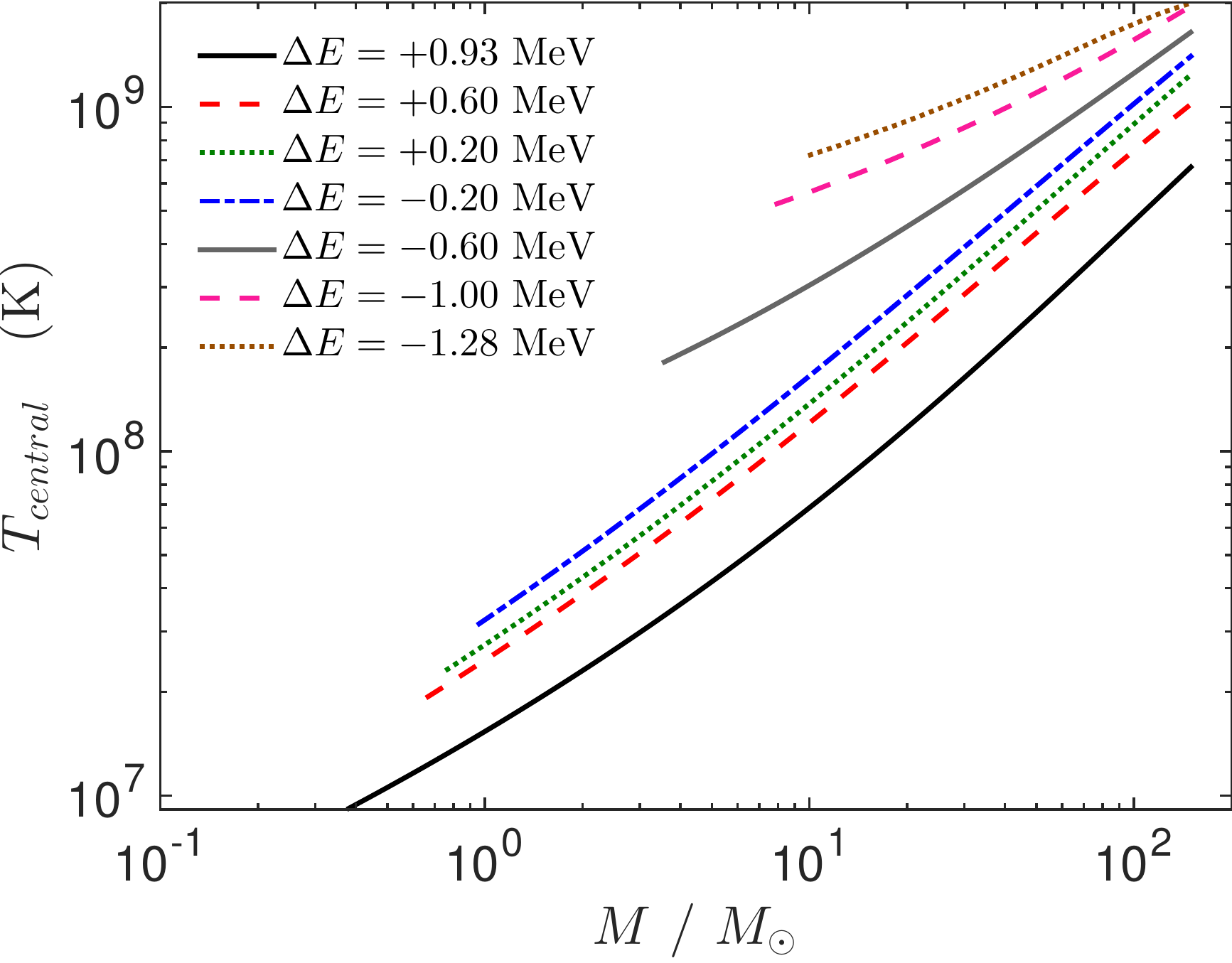}
	\end{minipage}
	\begin{minipage}{0.45\textwidth}
		\includegraphics[width=\textwidth]{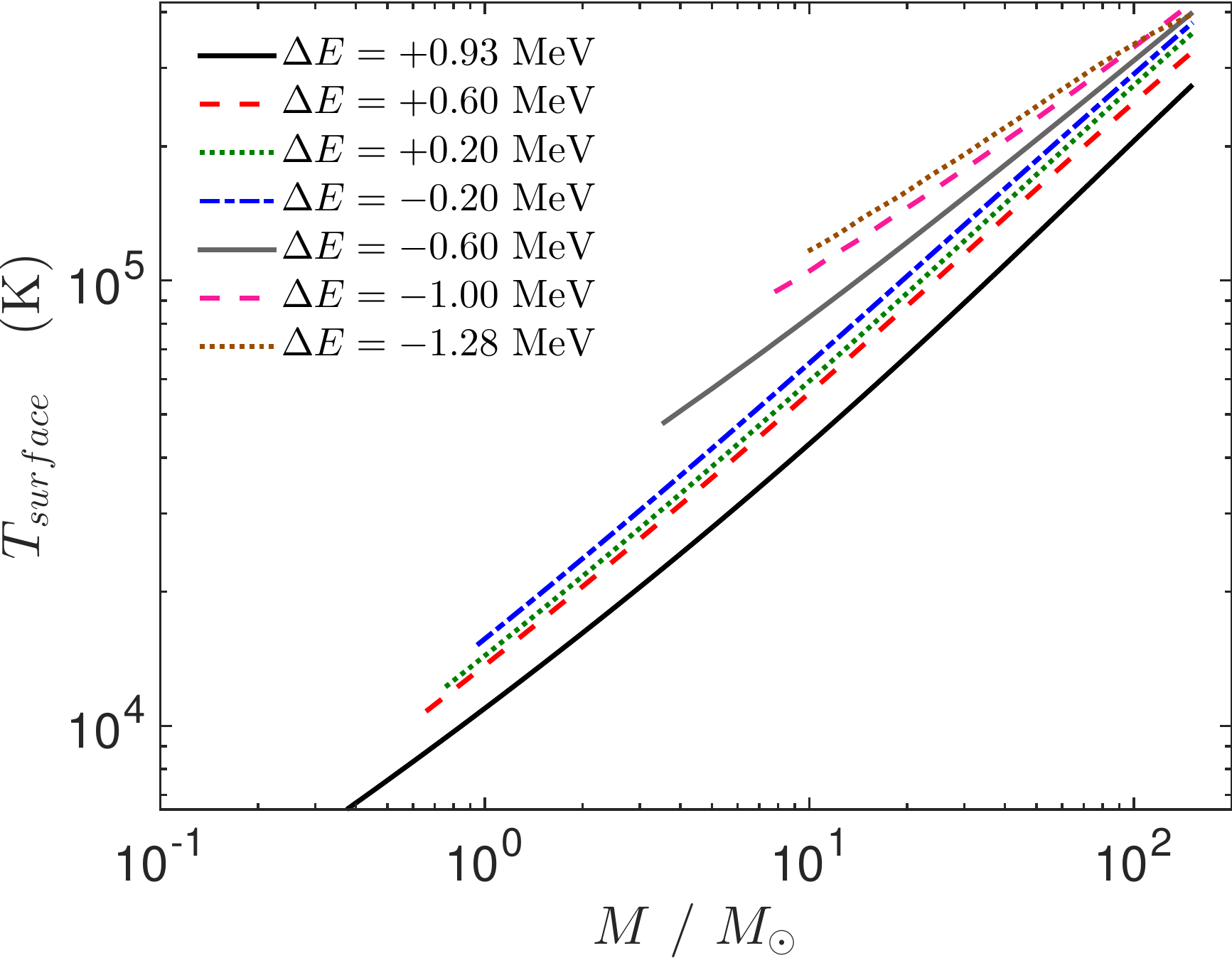}
	\end{minipage}
	\begin{minipage}{0.45\textwidth}
		\includegraphics[width=\textwidth]{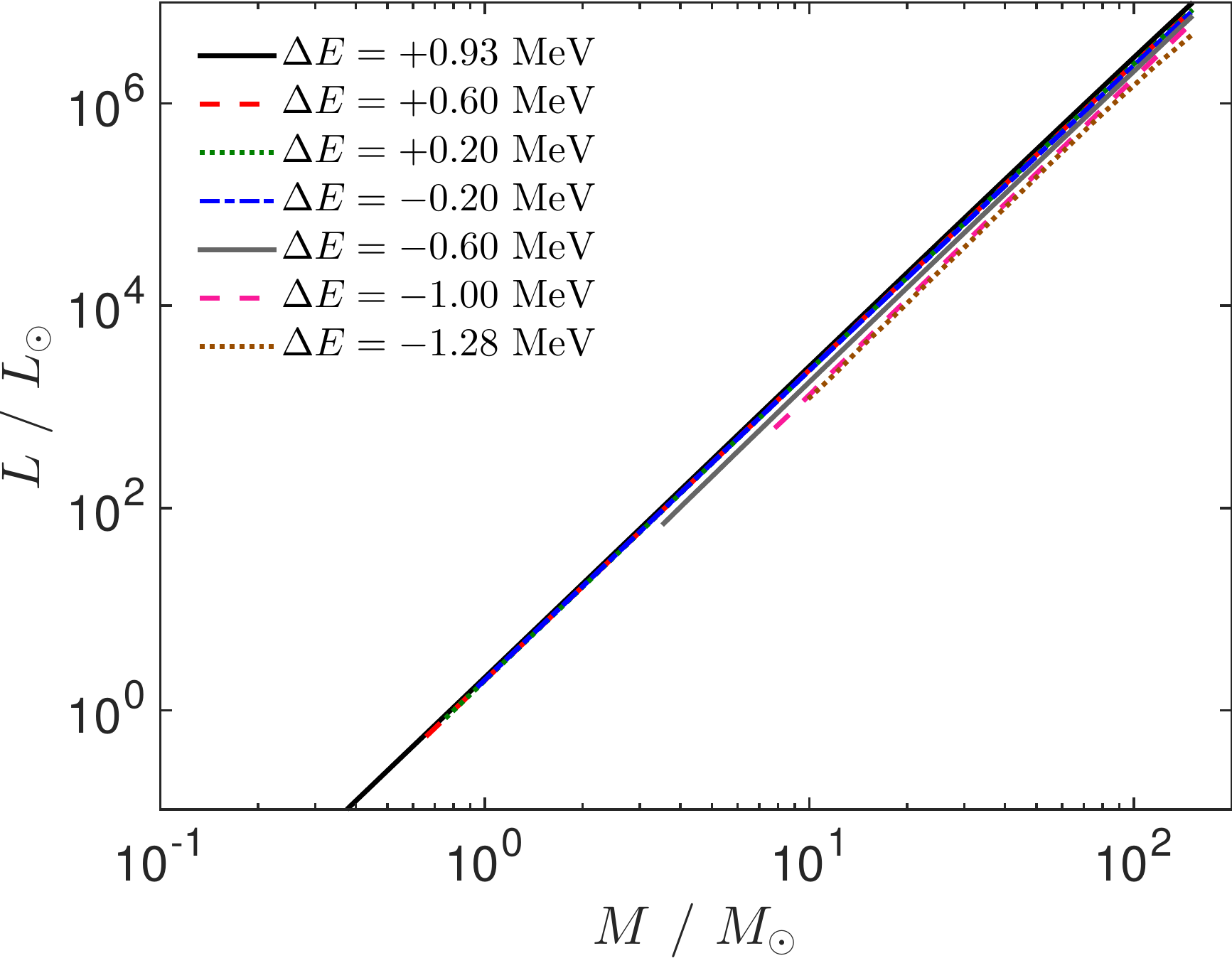}
	\end{minipage}
	\begin{minipage}{0.45\textwidth}
		\includegraphics[width=\textwidth]{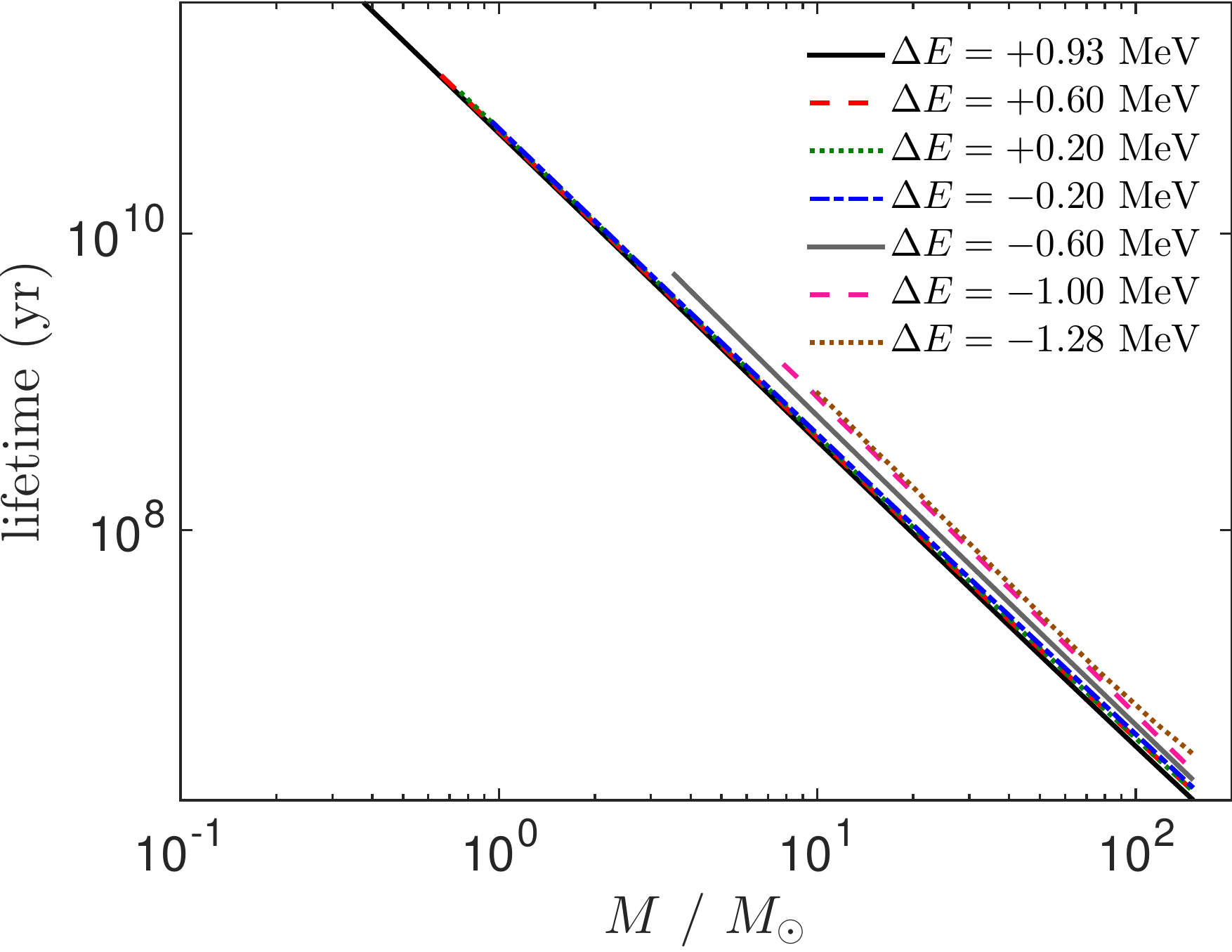}
	\end{minipage}
	\begin{minipage}{0.43\textwidth}
		\includegraphics[width=\textwidth]{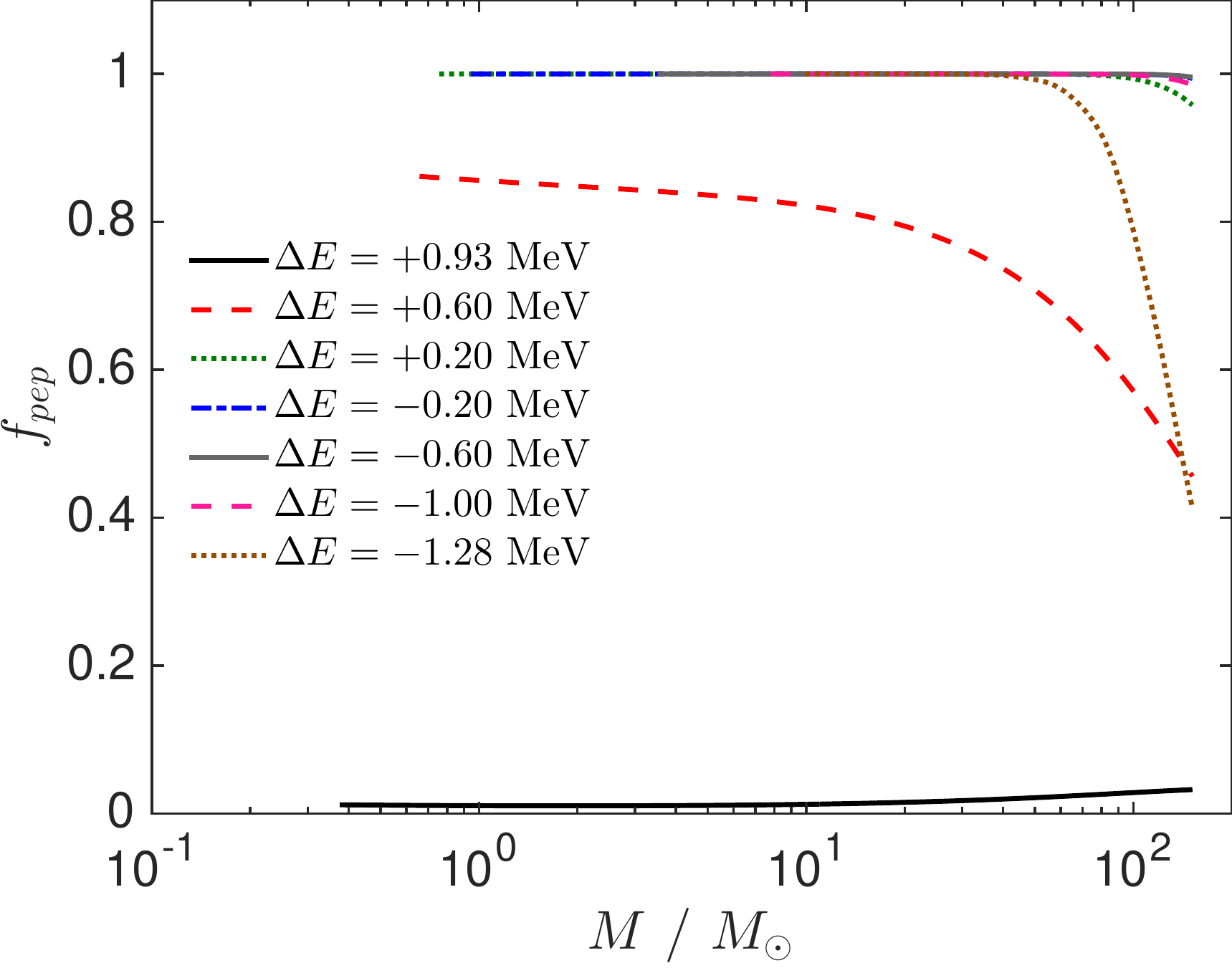}
	\end{minipage}
	\hspace{0.03\textwidth}
	\begin{minipage}{0.43\textwidth}
	\caption{Stars in universes in which the deuteron binding energy is reduced. The different colours of line show different values of the deuterium binding energy, corresponding to $\Delta E = B_D - (m_n - m_p) = (0.931,0.6,0.2,-0.2,-0.6,-1.0,-1.283)$ MeV. These plots highlight the effect of making the $pp$ and then also the $pep$ reactions endothermic --- small stars are increasingly unable to ignite nuclear reactions. The bottom right plot shows the fraction ($f_{pep}$) of deuterium producing reactions in the star that are via the $pep$ reaction.}
\end{minipage}
\label{fig:lessBound}
\end{figure*}

Figure \ref{fig:lessBound} shows the effect of reducing the deuteron binding energy. The different colours of line show different values of the deuterium binding energy. These plots highlight the effect of making the $pp$ and then also the $pep$ reactions endothermic. The top left plot shows the central temperature as a function of stellar mass. As expected, larger central temperatures are required at a fixed mass in order for the less-efficient $pp$ and $pep$ reactions to provide the luminosity to support the star. As a result, smaller stars are increasingly unable to ignite at all, so that the minimum stellar mass increases as the deuteron binding energy is decreased. The top right plot shows that the surface temperature similarly increases, with only massive, very hot stars able to burn.

As we saw previously, the stellar luminosity and lifetime at a given mass are largely unaffected. Note that the effect of neutrino cooling is largely negligible in these stars, because the deuteron lifetime is still relatively long. Even for a barely bound deuteron ($\Delta E  = -1.283$ MeV), the decay constant is $\lambda_D = 0.48$ s$^{-1}$. Thus, the major effect is the lack of smaller stars: the longest lived stars when the deuteron is barely bound are around 10 times shorter lived than the Sun.

The bottom right plot shows the fraction ($f_{pep}$) of deuterium producing reactions in the star that are via the $pep$ reaction. When the $pp$ reaction is endothermic  ($\Delta E < m_e = 0.511$ MeV), stars burn almost exclusively via the $pep$ reaction at most masses. 

In summary, unbinding the deuteron to the point that the $pp$ and $pep$ reaction are endothermic has a considerable effect on the properties of stars. The minimum stable stellar mass increases by a factor of about 100, so that only $10 - 100 \Msol $ stars can burn. These burn out more quickly than stars in our universe. Such universes would be less suitable for life, but are not definitively uninhabitable.

\section{Varying the Electron Mass} \label{S:varyme}

\begin{figure*} \centering
	\begin{minipage}{0.40\textwidth}
		\includegraphics[width=\textwidth]{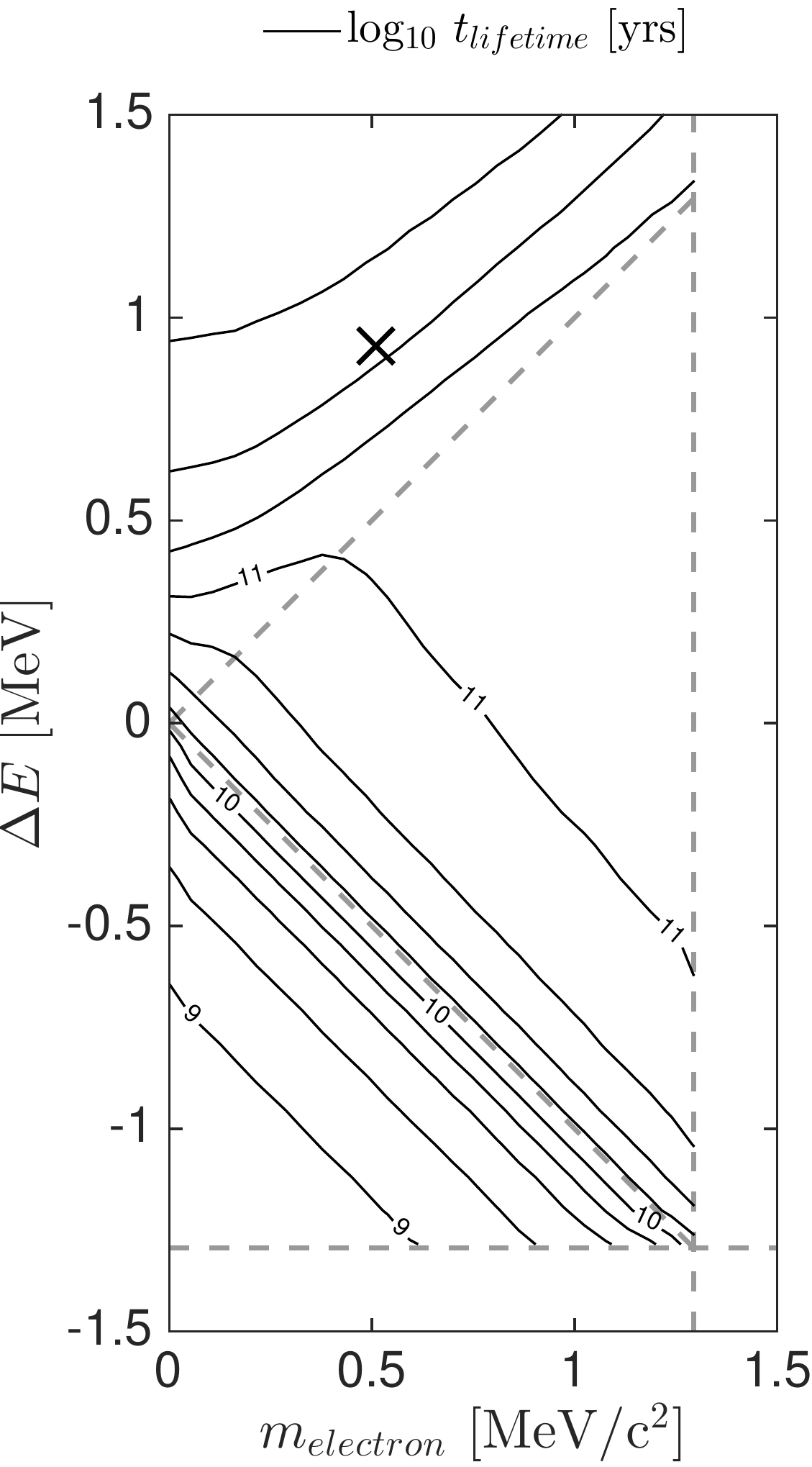}
	\end{minipage}
	\hspace{0.05\textwidth}
	\begin{minipage}{0.40\textwidth}
		\includegraphics[width=\textwidth]{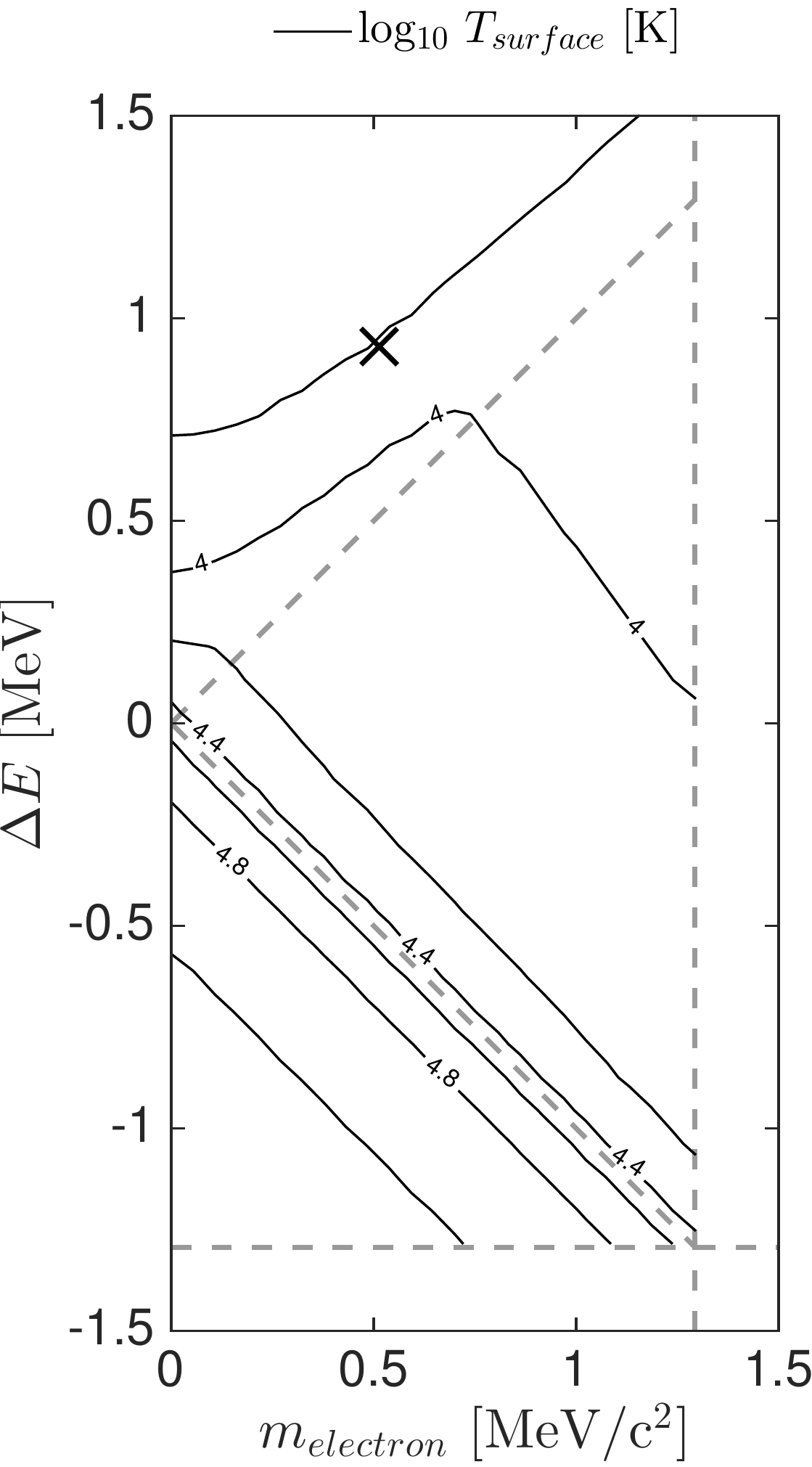}
	\end{minipage}
	\caption{The maximum stellar lifetime (left) and surface temperature (right) on the region of parameter space shown in Figure \ref{fig:Dme}; recall that $\Delta E = B_D - (m_n - m_p)$. Specifically, the contours on the left panel are labeled with the age of the longest lived (and hence, least massive) star possible in a universe with a given set of constants. The contours on the right panel are labeled with the surface temperature of the star whose temperature is closest to that of the Sun, which turns out to be the smallest stars for most of our models. Our Universe is shown as a black cross. The lower dashed grey line shows where the deuteron becomes unbound. The right dashed grey line shows where $m_p + m_e = m_n$. To the right of this line (holding $m_n - m_p$ constant), hydrogen is unstable to the reaction $p + e \rightarrow n$, and the universe becomes neutron dominated after recombination.}
\label{fig:varyelectron}
\end{figure*}

In this section, we will explore the full parameter space of Figure \ref{fig:Dme}, varying both $\Delta E$ and $m_e$. Rather than plotting stellar models as above, we will focus on two of the most relevant stellar properties for life: stellar lifetime, and stellar surface temperature. Figure \ref{fig:varyelectron} shows the maximum stellar lifetime (left) and surface temperature (right) on the region of parameter space shown in Figure \ref{fig:Dme}. Specifically, the contours on the left panel are labeled with the (log of the) age of the longest lived (and hence, least massive) star possible in a universe with a given set of constants. The contours on the right panel are labeled with the surface temperature of the star whose temperature is closest to that of the Sun, which turns out to be the smallest stars for most of our models. Our Universe is shown as a black cross. The lower dashed grey line shows where the deuteron becomes unbound. The right dashed grey line shows where $m_p + m_e = m_n$. To the right of this line (holding $m_n - m_p$ constant), hydrogen is unstable to the reaction $p + e \rightarrow n + \nu_e$, and the universe becomes neutron dominated after recombination \cite{Damour2008}.

We see that, when both the $pp$ and $pep$ reactions are endothermic, a smaller value of the electron's mass leads to shorter lived, hotter stars. Stars have temperatures of at least $\sim 10^5$ K, emitting photons with typical energies in excess of atomic ionization energies, which are very damaging to biological organisms. Looking over this small piece of parameter space, we see that it is possible to find long-lived ($\sim$ Gyr), perhaps life-sustaining (rather than life-sterilizing) stars. The boundaries in parameter space at which the $pp$ and $pep$ reactions become endothermic represent problems for living organisms, but do not mark the onset of a catastrophic failure of habitability. 

\section{Stars in Parameter Space} \label{S:starparam}

The most definitive boundary in parameter space that divides probably life-permitting universes from probably life-prohibiting ones is between a bound and unbound deuteron. Due to neutrino losses, balls of gas will undergo rapid cooling or stabilization by electron degeneracy pressure before they can form a stable, nuclear burning star. By contrast, the transition to endothermic $pp$ and $pep$ reactions, and the resulting beta-decay instability of the deuteron, do not seem to represent catastrophic problems for life.

We repeat our \emph{caveat} from the introduction: in light of the uncertain relation between the fundamental masses of elementary particles --- particularly the quark masses and electron mass --- and the properties of the deuteron, we have varied the deuteron binding energy by varying the strong force coupling constant only. A $\sim 5-8$\% decrease in the strength of the strong force unbinds the deuteron \cite{Davies1972,Davies1983,Pochet1991,Cohen2008,Golowich2008}.
A wider look at parameter space, varying all the fundamental constants including the quark masses, is left for future work.

As this manuscript was in preparation, Adams and Grohs \cite{Adams2016} published an investigation of stars in universes in which the deuteron is unstable. We have shown here that such stars will not ignite nuclear reactions. Adams and Grohs look more closely at the evolution of protostars in such universes, noting that they can still generate energy by gravitational contraction. Somewhat surprisingly, there can be an extended period of almost-constant luminosity from ``stars'' powered only in this way. Here, we have also considered the effect of a less-stable deuteron. Their results are complementary to ours.

\acknowledgments
LAB: For Grandad. Supported by a grant from the John Templeton Foundation. This publication was made possible through the support of a grant from the John Templeton Foundation. The opinions expressed in this publication are those of the author and do not necessarily reflect the views of the John Templeton Foundation. Thanks also to Fred Adams and Evan Grohs for their useful comments.


\end{document}